\documentclass[floatfix,prl,showpacs,twocolumn,preprintnumbers,amsmath,amssymb]{revtex4}
\usepackage{graphicx}
\def\pmb#1{\setbox0=\hbox{#1}
\kern-.025em\copy0\kern-\wd0
\kern.05em\copy0\kern-\wd0
\kern-.025em\raise.0433em\box0}

\begin{document}

\title{Spurious fixed points in frustrated  magnets}

\author{B. Delamotte$^{1}$, Yu. Holovatch$^{2,3}$, D. Ivaneyko$^{4}$, D. Mouhanna$^{1}$
 and M. Tissier$^{1}$}

\address{$^{1}$ LPTMC, CNRS-UMR 7600, Universit\'e Pierre
et Marie Curie, 75252 Paris C\'edex 05, France}

\address{$^{2}$
Institute   for  Condensed  Matter Physics  and  Ivan  Franko National
University of Lviv, UA--79005 Lviv, Ukraine}

\address{$^{3}$
Institute f\"ur Theoretische Physik,  Johannes Kepler  Universit\"at Linz,
A-4040 Linz, Austria}

\address{$^{4}$
Ivan Franko National University of Lviv, UA--79005 Lviv, Ukraine}

\begin{abstract}

We analyze  the validity of   perturbative  estimations obtained 
at fixed  dimensions  in the study  of frustrated magnets.  To
this end we  consider the five-loop  $\beta$-functions obtained within the
minimal subtraction   scheme and exploited without  $\epsilon$-expansion both
for   frustrated  magnets and  for the   well-controlled ferromagnetic
systems with a  cubic anisotropy.  Comparing  the two cases it appears
that the fixed point supposed     to control the second order    phase
transition   of  frustrated  magnets is    very  likely an  unphysical
one. This is  supported by  the non-Gaussian  character of this  fixed
point  at the upper  critical dimension $d=4$.   Our work confirms the
weak first order nature of the phase transition and constitutes a step
towards a unified picture of existing   theoretical approaches to frustrated
magnets.

\end{abstract}
\pacs{75.10.Hk, 11.10.Hi, 12.38.Cy}

\maketitle


\vskip 1cm

{\sl Introduction.}  Although  undoubtedly successful to describe  the
critical behavior of $O(N)$-like  models, perturbative field theory is
still unable to provide   a clear, non-controversial  understanding of
the   physics  of more  complex  models  among  which  are the  famous
Heisenberg  or $XY$   frustrated magnets (see  \cite{delamotte03}  and
references therein).   At the  core of the  problem  is that different
kinds  of perturbative approaches, performed up  to  five- or six-loop
order,  lead to contradictory  results:  first order phase transitions
are predicted within the $\epsilon$ (or pseudo-$\epsilon$)- expansion
\cite{antonenko95,holovatch04,calabrese03c} whereas a second order
transition  is  found   in  the  fixed-dimension  (FD)    perturbative
approaches   performed      either    in  the      minimal-subtraction
($\overline{\hbox{MS}}$) scheme {\sl without} $\varepsilon$-expansion
\cite{calabrese04} or in the massive scheme \cite{pelissetto01a}. 
 In fact  FD results for frustrated magnets  are neither  supported by
 experiments nor by Monte Carlo simulations
\cite{itakura01,peles04,bekhechi06,quirion06} (see however
\cite{calabrese04} where a scaling behavior is found).  They
also disagree with the predictions  obtained from the non-perturbative
renormalization   group (NPRG) approach   \cite{tissier00,delamotte03}
that predicts (weak) first order phase transitions in $d=3$ in agreement with
the $\epsilon$-expansion analysis.

In this letter we  shed   light on  the discrepancies encountered   in
perturbative  approaches to frustrated  magnets by showing that the FD
approaches very likely  lead    to incorrect predictions  as  for  the
critical physics in three dimensions.   Our key-point is easy to grasp
already for the simplest  ---  $O(N)$ ---  model.  In this  case,  the
(non-resummed) renormalization group (RG) $\beta$-function at $L$ loops is
a polynomial in  its coupling constant $u$  of  order $L+1$. Thus,  it
admits $L+1$ roots $u^*$, $\beta(u^*)=0$, that are either real or complex.
Within the $\varepsilon$-expansion   approach, the only non-trivial  fixed point
(FP)  retained is such that  $u^*\sim \epsilon$ where $\epsilon=4-d$.  On the contrary,
in  the FD approaches, no real  root can be  {\sl a priori} discarded.
As a result, the generic  situation is that  the number of FPs as well
as their stability vary  with the order $L$:  at a given  order, there
can exist several real and stable FPs or none instead of a single one.
This  artefact of  the  FD approach  is  already  known and was  first
noticed in  the massive scheme in $d=3$   \cite{parisi80}.  The way to
cope  with it is also known:   resumming the perturbative expansion of
$\beta(u)$ (see  e.g.  \cite{zinnjustin89,pelissetto01c}) is supposed both
to  restore  the nontrivial     Wilson-Fisher  FP and  to suppress    the
nonphysical or ``spurious'' roots.  This is indeed what occurs for the
$O(N)$  model for which  the  FP analysis  performed  on the  resummed
$\beta$-function of FD approaches enables to discriminate between physical
and  ``spurious'' FPs.  We argue that  the situation is very different
for frustrated magnets.  Indeed, considering the $\beta$-functions derived
at five loops in the $\overline{\hbox{MS}}$  scheme and using the same
resummation procedure as Calabrese {\it  et al.} \cite{calabrese04} we
show that the  FP found in $d=3$  without expanding in $\epsilon$ is spurious
although it  persists after resummation.   Our conclusion  is based on
two main facts: (i) analyzed with the same FD approach a similar FP is
found for the    ferromagnetic   system  with  cubic   anisotropy   in
contradiction  with its well   established critical physics (ii) there
are strong  indications that these  FPs survive in the  upper critical
dimension $d=4$ where they are  found to be non-Gaussian which  raises
serious doubts on their actual existence.

 {\sl  Resummation.}    To investigate   the  five-loop $\beta$  functions
 derived in the $\overline{\hbox{MS}}$ scheme we resum them, following
 \cite{calabrese04},  using   a  conformal   mapping Borel   transform
 suitable for series involving two coupling constants $u_1$ and $u_2$.
 We consider any   function $f$ of $u_1$   and $u_2$ as  a function of
 $u_1$  and $z=u_2/u_1$,   supposed  to be   known through its  series
 expansion:
\begin{equation}
f(u_1,z)=\sum_{n} f_n(z) \ u_1^n \ .
\label{series}
\end{equation}
An important hypothesis underlying the  procedure of resummation  used
here  and   in \cite{calabrese04}  is  that    one can  safely   resum
(\ref{series}) in the $u_1$ direction   while keeping $z$ fixed,  {\it
i.e.} without resumming with  respect to $u_2$.  Under this hypothesis
a resummed expression associated with $f$ reads:
\begin{equation}
f_R(u_1,z)=\sum_{n} d_n(\alpha,a(z),b;z) \hspace{-0.1cm} \int_0^{\infty}
\hspace{-0.2cm}dt\, \, {e^{-t}\, t^{b} \left[\omega(u_1 t;z)\right]^n    \over 
\left[1-\omega(u_1 t;z)\right]^{\alpha} }
\label{resummation}
\end{equation}
with $\omega(u;z)={(\sqrt{1  + a(z)\, u}-1)/ (\sqrt{1  + a(z)\, u}+1})$ and
where the  coefficients   $d_n(\alpha,a(z),b;z)$ are computed   so that the
re-expansion of the r.h.s.  of (\ref{resummation})  in powers of $u_1$
coincides  with that of  (\ref{series}).   In principle the parameters
$a(z)$, $b$ and $\alpha$  are determined (i)  by the asymptotic behavior of
$f_n$: $f_{n\to\infty}\sim   (-a(z))^n  \, n!\, n^b$ and    (ii) by  the  strong
coupling  behavior  of  $f$:  $f(u_1\to\infty,z) \sim u_1^{\alpha/2}$.     However in
general only  $a$ is known and the  other parameters are considered as
free or variational.

\begin{figure}[htbp] 
\vspace{0cm}
\hspace{3cm}
\includegraphics[width=0.8\linewidth,origin=tl]{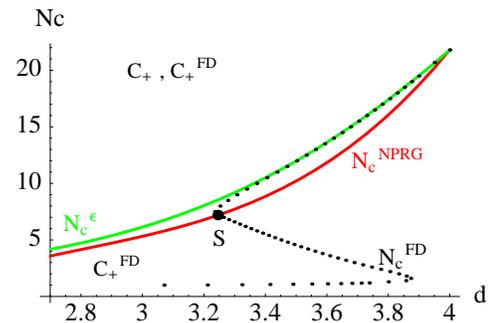}\hfill%
\caption{Curves $N_c(d)$ obtained within the $\epsilon$-expansion ($N_c^\epsilon$), the  $\overline{\hbox{MS}}$
 scheme  without $\epsilon$-expansion ($N_c^{\text{FD}}$) and the NPRG approach ($N_c^{\text{NPRG}}$). 
The resummation parameters for the $\overline{\hbox{MS}}$ curve are $a=1/2$, $b=10$ 
and $\alpha=1$. The  part of the  curve $N_c^{\text{FD}}$ below S  corresponds
 to a regime of non-Borel-summability.} 
\label{courbes_ncd} 
\end{figure}

 {\sl  Frustrated magnets.}   The Hamiltonian  relevant for frustrated
 systems is given by:
\begin{equation}
\begin{array}{ll}
\displaystyle \hspace{0cm}{\mathcal H}= \int{\rm d^d} x \Big\{\frac{1}{2}
\left[(\partial\pmb{$\phi$}_1)^2+
 (\partial\pmb{$\phi$}_2)^2 + m^2 (\pmb{$\phi$}_1^2+\pmb{$\phi$}_2^2)\right]+\\
\\
\hspace{1cm}\displaystyle \frac{u_1}{4!}[\pmb{$\phi$}_1^2+\pmb{$\phi$}_2^2]^2 
+\frac{u_2}{12}[(\pmb{$\phi$}_1 \cdot \pmb{$\phi$}_2)^2-
\pmb{$\phi$}_1^2\pmb{$\phi$}_2^2 ] \Big \}
\end{array}
\label{landau}
\end{equation}
where the $\pmb{$\phi$}_i$ are  $N$-component vector fields. We apply the
resummation procedure described above without $\epsilon$-expansion
\cite{schloms87} to the $\beta_{u_i}$ functions, $i=1,2$, obtained at five
loops  in the $\overline{\hbox{MS}}$  scheme  \cite{calabrese04}. More
precisely, as in
\cite{calabrese04}, we  resum  $(\beta_{u_i}(u_1,z) + \varepsilon u_i)/{u_1}^2$,
$i=1,2$, instead of $\beta_{u_i}(u_1,z)$ \footnote{Resumming the functions
$\beta_{u_i}$ leads  to similar  results.}.  In  this  model the region of
Borel-summability is given by  $u_1-u_2/2>0$ while $a(z)=1/2$  and $b$
and $\alpha$  are typically varied  in the ranges $[6,16]$  and $[-0.5,2]$.
As in all  other       approaches, one finds, in    agreement     with
\cite{calabrese04},  that  there  exists  a  curve   (parametrized  by
$N_c(d)$ or its reciprocal $d_c(N)$) such that for $d<d_c(N)$ a stable
FP $C_+$  governs the critical properties of  the system.   The curves
$N_c(d)$  obtained  within  this scheme,   the  NPRG approach  and the
$\epsilon$-expansion are given in Fig.\ref{courbes_ncd}.  This figure sums up
the  discrepancies between these  approaches.  The S-like shape of the
curve  $N_c^{\text{FD}}(d)$  obtained   within the   perturbative   FD
approach  is such that  $C_+^{\text{  FD}}$ exists  for  all $N\geq 2$ in
$d=3$ contrary to the other approaches in which a FP $C_+$ exists only
for $N>N_c(d=3)\simeq5$ \cite{delamotte03}.

\begin{figure}[htbp] 
\vspace{-0cm}
\hspace{3cm}
\includegraphics[width=0.7\linewidth,origin=tl]{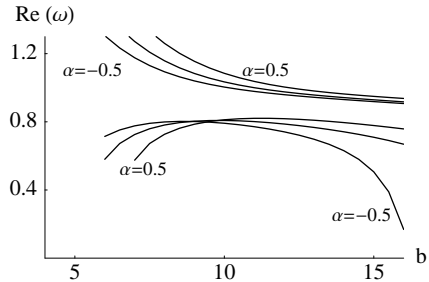}\hfill%
\caption{The (real part of the) critical exponent $\omega$ as a function of $b$ at five (upper curves)
 and  four (lower curves)  loops for $\alpha=-0.5,0$ and  0.5  for the
 frustrated model ($N=3$).}
\label{frustre} 
\end{figure}

The   main question  is  to know   whether  one can trust  the results
obtained  at FD, in  particular  the  existence  of  FPs in $d=3$  for
$N=2,3$.  To answer this question we examine the convergence of the FD
perturbative  results  through   their sensitivity   with  respect  to
variations of  the resummation parameters  $b$ and $\alpha$  as well as the
order $L$ of the  computation. We concentrate our  study on the  (real
part of the) stability critical exponent $\omega$ at the  FP $C^+$ which is
focus
\cite{calabrese02}.  We have checked that similar behaviors are
obtained  for  the exponent   $\nu$ \cite{delamottenext}.   In practice,
following
\cite{mudrov98c}, we optimize $\omega(\alpha,b,L)$ by choosing $\alpha$ such that 
$\omega(L+1)-\omega(L)$     is minimal and    $b$      such  that $\omega(b)$      is
(quasi-)stationary. In   Fig.\ref{frustre}  we display   $\omega(b)$ in the
Heisenberg case for $L=4$ and 5 for typical values of $\alpha$.  At a given
order  the variations of $\omega$ with  $b$ are small around the stationary
point  and do not exceed 30$\%$  from four  to  five loops in agreement
with  \cite{calabrese04}. Similar  results   are  obtained in  the  XY
case.  From this analysis alone,  one could conclude  that a FP indeed
exists in $d=3$ although with error bars on the critical exponents far
larger  than  those  obtained for    the Ising  model  with  the  same
methodology  (much less  than 1$\%$ at this  order).  Let us now
study the cubic model along the same lines.

 {\sl  The cubic model.} We  now consider the ferromagnetic model with
 cubic anisotropy whose Hamiltonian is:
\begin{equation}
\displaystyle \hspace{0cm}{\mathcal H}= \int{\rm d^d} x \Big\{\frac{1}{2}\left[(\partial\pmb{$\phi$})^2+
m^2 \pmb{$\phi$}^2\right]+   {u\over     4!}   \left[\pmb{$\phi$}^2\right]^2
+{v\over 4!}\sum_{i=1}^N \phi_i^4\Big \}
\label{cubic}
\end{equation}
with  $\pmb{$\phi$}$  a   $N$-component vector   field.   The Hamiltonian
(\ref{cubic}) is  used to study  the  critical behaviors  of  numerous
magnetic  and ferroelectric  systems  with appropriate order parameter
symmetry (see e.g.  \cite{folk00b,pelissetto01c}).     The
$\beta$-functions   are    known     at    five-loop     order  in     the
$\overline{\hbox{MS}}$ scheme \cite{kleinert95}  and at six-loop order
in the massive scheme \cite{carmona00}. Apart from the Gaussian and an
Ising FP,  there  exist two   FPs:  the $O(N)$ symmetric FP   $(u^*\neq0,
v^*=0)$ and  the mixed  one $M$ $(u^*\neq  0,  v^*\neq 0)$.   The  Ising and
Gaussian FPs are both unstable  for all values of  $N$.  The $O(N)$ FP
is stable  and $M$ is unstable with  $ v^*< 0$ for $N<\tilde{N_c}$ and
the opposite for $N>\tilde{N_c}$.   $\tilde{N_c}$ has been found to be
slightly less than 3 \cite{carmona00,folk00b}.

Let us now analyze the FP structure of  the model (\ref{cubic}) within
the  $\overline{\hbox{MS}}$ scheme  without $\epsilon$-expansion by  applying
the  conformal  mapping  Borel   transform (\ref{resummation}).    The
parameter  $a(z=v/u)$ entering in  (\ref{resummation}) is now given by
$a(z)=1+z$ for  $z>0$ and $a(z)=1+z/N$ for  $z<0$ while  the region of
Borel-summability  is  given by the   condition  $u+v>0$ and $Nu+v>0$.
Within this scheme one surprisingly observes that,  in addition to the
above mentioned usual  FPs, there exist, in a  whole domain of parameters $b$
and   $\alpha$,  several  other  FPs   that   have  no counterpart   in the
$\epsilon$-expansion.  In particular, one of them that we  call $P$ (which is
stable and such that $u^*>0, v^*<0$) exists  for any value of  $N\lesssim
7.5$  and lies in the region   of  Borel-summability $u+v>0$.  The
presence of this FP, if taken seriously, would have important physical
consequences  since  it  would  correspond   to  a second  order phase
transition with a new universality  class. However as  far as we know,
no such transition has ever been reported and the existence of $P$ has
to be considered as an artefact of the FD  analysis. Note finally that
$P$ is found  to be a  {\sl focus} FP, a  striking similarity with the
frustrated case.

\begin{figure}[htbp] 
\vspace{-0cm}
\hspace{3cm}
\includegraphics[width=0.7\linewidth,origin=tl]{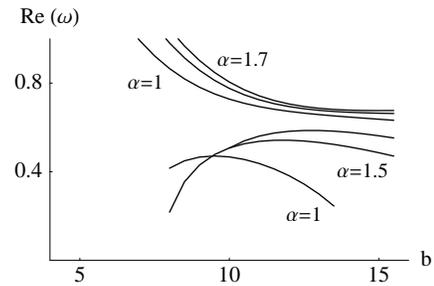}\hfill%
\caption{The (real part of the) critical exponent $\omega$ as a function of $b$ at five (upper curves)
 and four  (lower curves)  loops for  $\alpha=1,1.5$ and 1.7  for the
 cubic model (N=2).}
\label{omegacubic} 
\end{figure}

At this   stage, it is very  instructive  to perform for $P$  the same
convergence analysis as we have done for $C_+$ in  the FD analysis. In
Fig.\ref{omegacubic}   we  give  $\omega(b)$   for  $L=4,5$ and   for three
different values of  $\alpha$.  Surprisingly, we  find results that display
similar  convergence  properties as  in  the frustrated   case (with a
spreading of the  results of order $40\%$).

 This  study shows that
contrary to what one could naively  deduce, the convergence properties
displayed in Fig.\ref{omegacubic} do not characterize $P$ as a genuine
FP. Comparing the studies of the frustrated and cubic cases one is led
to  the  same conclusion  for $C_+^{\text{FD}}$ for  $N=2,3$ in $d=3$.
Thus, in order to conclude  whether $C_+^{\text{FD}}$  is a genuine  FP or an
 artefact of the FD analysis, one has to resort to another criterion.

\vspace{0.5cm}
{\sl Fixed point at the upper critical dimension.}  To  do this let us
perform   the  following    observation:  the   choice of   $\phi^4$-like
Hamiltonian in Eq.(\ref{landau})   relies on  the physical  hypothesis
that the upper critical dimension    is four.  From the   perturbative
point  of view, this means in  particular  that the theory is infrared
trivial  ---     controlled  by the    Gaussian   FP    ---   in  four
dimensions. Thus, by consistency, one  must retain among all solutions
of the  FP equations in  $d=4-\epsilon$ only those that are  at a distance of
order $\epsilon$ of the  Gaussian  FP.  Therefore a  practical way  to  check
whether a FP   found at a  given  dimension, $d=3$ for  instance, is a
genuine FP or is just an artefact of perturbation  theory is to follow
it  by continuity up   to the upper    critical dimension \cite{dudka04,holovatch04}.
  If  the FP
survives as a non-Gaussian FP at this  dimension this is a signal that
it is spurious.

\begin{figure}[htbp] 
\begin{center}
\includegraphics[width=0.6\linewidth]{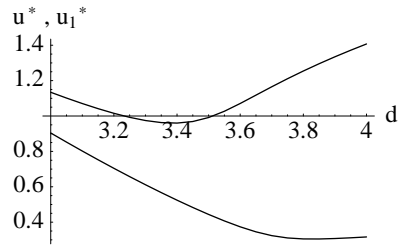}
\end{center}
\caption{The $u^*$ coordinate of the FP $P$ ($N=2$, upper curve) and the 
$u_1^*$ coordinate of the FP $C_+^{\text{FD}}$ ($N=3$, lower curve)  as  functions  of $d$.} 
\label{frustre3} 
\end{figure}

Let    us  apply  this   criterion     simultaneously   to   $P$   and
$C_+^{\text{FD}}$.  We present our results in Fig.\ref{frustre3} where
we have   displayed the coordinates $u^*$    and $u_1^*$ associated to
these  FPs as   functions of  $d$.   Manifestly,  they  both  survive
everywhere above $d=3$  and are {\sl  not} Gaussian in $d=4$.  In  the
cubic case this is true for all values of $N$ for which $P$ exists 
while, in the frustrated
case, this happens  for all values of  $N\lesssim 6$.  Thus, according to our
criterion,  $P$ is, as expected,  always found to  be spurious. As for
$C_+^{\text{FD}}$  for   $N\lesssim6$  and, in particular,    for $N=2,3$ our
criterion strongly   suggests   that it is   also    spurious in $d=3$
\footnote{Let  us  clarify   a  point   specific  to  the   frustrated
systems.  The curve $N_c^{\text{FD}}(d)$    can be separated  into two
parts by  the point  $S$,  see Fig.\ref{courbes_ncd}.   If one follows
$C_+^{\text{FD}}$ along  a path starting  in $d=3$, going to $d=4$ and
crossing $N_c^{\text{FD}}(d)$  above   $S$,  its  coordinates   become
complex  in $d=d_c(N)$ and go  to zero for  $d=4$ where it is thus the
Gaussian FP.       If,     on  the contrary,   the      path   crosses
$N_c^{\text{FD}}(d)$ below  $S$,  $C_+^{\text{FD}}$  changes   from  a
stable focus to an unstable  one at $d=d_c(N)$ and  does not go to the
Gaussian  in $d=4$.  Thus,  $S$ is a  singularity of  $u_1^*(N,d)$ and
$u_2^*(N,d)$.   The FD results  are  therefore doubtful below $S$  but
trustable above where they are indeed found  to be consistent with the
other approaches, see Fig.\ref{courbes_ncd}. }.  Note that the FPs $P$
and $C_+^{\text{FD}}$  lie out of the  region of  Borel-summability in
$d=4$. Thus their coordinates cannot be determined accurately although
their existence as non-Gaussian FPs is doubtless.

\begin{figure}[tbp] 
\vspace{0cm}
\hspace{3cm}
\includegraphics[width=0.8\linewidth,origin=tl]{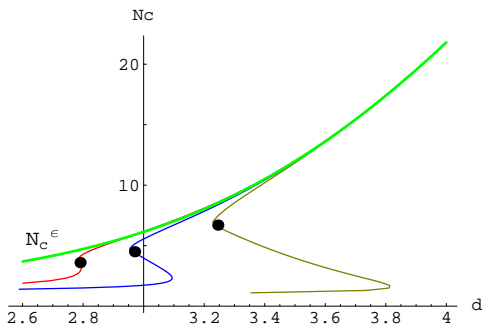}\hfill%
\caption{Four curves $N_c^{\text{FD}}(d)$ for different values of the parameter $a$ 
and the curve $N_c^{\epsilon}(d)$. From right to left $a=0.5, 1.3, 1.5$. The
 parameters $b$ and $\alpha$ are $b=10$ and $\alpha=1$. The parts of the curves
 below the black dots  correspond to a regime of non-Borel summability. } 
\label{nc} 
\end{figure}

{\sl   Conclusion}.   It   appears   from our   study  that   the   FP
$C_+^{\text{FD}}$  identified in   the   FD approach  is  very  likely
spurious.   The transition should  thus  be of  --- possibly weak  ---
first order  in agreement with  NPRG and $\epsilon$-expansion approaches.  It
remains to  explain the failure in the  resummation procedure  used in
the FD  approach, Eq.(\ref{resummation}).  As already  emphasized this
procedure  relies  on the  hypothesis  that resumming with  respect to
$u_1$,  keeping a polynomial structure  in $u_2$, is sufficient.  
Alternatively  a resummation of the series with respect to the {\sl two}
coupling constants could  be required to  obtain reliable results (see
for instance  \cite{alvarez00}   for  the   randomly  diluted    Ising
model).  Postponing  these considerations for a future publication
\cite{delamottenext} we assume that  the use of Eq.(\ref{resummation})
as such is  justified.  Then a possible  origin of the  failure in the
resommation  procedure could be  that  the  series considered are  not
known at large enough order to reach the asymptotic behavior.  In this
case  there  would be   no reason  to  fix  the  parameter $a$  at its
asymptotic $1/2$ value and one  should have to vary it  as $b$ and $\alpha$
to optimize the  results \cite{mudrov98c}. We display in  Fig.\ref{nc}
the curves $N_c^{\text{FD}}(d)$ for different values of $a$.  The part
corresponding to  large values  of  $N_c$, typically  for  $N_c\gtrsim6$, is
almost  insensitive to the  variations of $a$  whereas this is clearly
not the  case  for  smaller  values  of   $N_c$.  In particular,   for
sufficiently large  values of $a$, typically  $a\geq1.5$, the S-like part
is pushed below  $d=3$  so that $N_c^\epsilon(d)$,   $N_c^{\text{FD}}(d)$ and
$N_c^{\text{NPRG}}(d)$  are  then  compatible everywhere for  $3<d<4$.
Thus, under our hypothesis,   all qualitative differences  between the
different  approaches  disappear \footnote{One also  observes that the
shape of the curve $N_c^{\text{FD}}(d)$ for $a=1.3$ is compatible with
the results obtained in the massive scheme in $d=3$ in which one finds
fixed points for all values one $N$ but in the range $5.7(3)<N<6.4(4)$
\cite{calabrese03b}.}  so that  the  problem  would  boil  down to   a
question  of  order of computation.   It  is clear that  our heuristic
argument, although appealing, requires more rigourous developments
\cite{delamottenext}.

Note  finally that our  present  considerations surely pertains to the
case of frustrated magnets in $d=2$ \cite{calabrese02}. Indeed we have
checked that the FP  found in $d=2$ is continuously  related to the FP
$C_+^{\text{FD}}$ in $d=3$ which  makes  its existence doubtful.   Our
conclusions could also  apply in other  situations where FPs that have
no counterpart in the  $\epsilon$-expansion is found, as it  is the case, for
instance, in QCD at finite temperature \cite{basile04}.

\begin{acknowledgments} 
 We wish to thank P. Azaria, P. Calabrese, R. Guida and J. Zinn-Justin
 for useful discussions.
\end{acknowledgments}

\end{document}